\documentclass[prb,preprint]{revtex4-1} 

\usepackage{amsmath}  
\usepackage{amsfonts} 
\usepackage{graphicx} 

\begin{document}


\title{Cold atoms interacting with highly twisted laser beams mimic the forces involved in Millikan's experiment}

\author{Vasileios E. Lembessis}
\email{vlempesis@ksu.edu.sa} 
\affiliation{Quantum Technology Group, Department of Physics and Astronomy, College of Science, King Saud University, P.O. Box 2455, Riyadh 11451, Saudi Arabia}

\date{\today}

\begin{abstract}
This paper considers the analogy between the force exerted on cold atoms when they interact with a highly twisted tightly focused laser beam and the forces exerted on a charged dielectric particle inside a uniform electric field when we perform the Millikan's experiment. In the later case the particle experiences its weight, a force due to the electric field which is arranged in the opposite direction of the weight,  and a drag force due to the air proportional and opposite to the velocity of the particle. The force due to the electric field is "quantised" since the charge
of the particle is always an integer multiple of the electron charge. In the case of the cold atoms the total force is made up by three terms. The second term is quantised since it is proportional to the beam helicity which is an integer number. The sign of this term is opposite to the first. The third term is a damping force force proportional and opposite to the velocity of the atom.
We present numerical calculations with parameters taken from experimental data which show an impressive analogy between the involved forces.  

\end{abstract}

\maketitle 

\section{Introduction} 

It is not an exaggeration to say that in the last four decades we have witnessed  
the renaissance of atomic physics. The advent of tunable lasers gave us the opportunity 
to interact resonantly with atoms transferring initially momentum and recently angular momentum in their gross motion. 
As a result we managed from the one hand to cool the atomic motion and from the other to trap it
in specific regions of space. The cold atoms, thus, gave us new exciting opportunities which up to then were
remaining at the level of thought experiments. Among the most prominent applications of cooling and trapping of the atomic motion we refer to the possibility of Doppler free spectroscopy, the creation of optical molasses, 
the optimisation of the operation and quality of atomic clocks and the achievement of the the Bose-Einstein condensation in dilute atomic samples.\cite{cto11} Moreover cooling and trapping constitute sine qua non conditions for  the experiments on single atoms and ions which gave us the opportunity to test fundamental
principles of quantum mechanics and gave new horizons in quantum computation.\cite{hr06} 
Finally atom cooling and trapping are inherently associated with the fresh air which blows over condensed matter
physics with the creation of optical lattices which enabled the simulation of condensed matter physics effects and the creation of artificial gauge magnetic and electric fields.\cite{lsa12}

The physical basis of atom cooling is the continuous removal of photons from a laser beam when it interacts with an atom.
Since the laser beam has a sharply defined frequency it induces transitions of an atomic electron
between two atomic levels, one with lower energy called the ground state and another with higher energy called the excited state.
Each photon of the laser beam carries a momentum equal to $\hbar k$, where $k=2\pi/\lambda$ with $\lambda$ being the laser light wavelength. 
The atom can absorb a photon of the beam and transferred to the excited state while simultaneously
its gross motion acquires a momentum equal to $\hbar k$ along the beam propagation direction. 
Being in the excited state the atom could emit a photon back to the beam (stimulated emission) and return to its 
ground state while its gross motion gets a momentum "kick" equal to $\hbar k$ in the direction opposite to the beam's propagation
 direction. It is obvious that such a cycle of photon absorption-emission does not change the atomic momentum. 
 But this is not the whole story as the atom has another way to emit a photon while being in its excited state. 
This is spontaneous photon emission which results from the fluctuations of the electromagnetic vacuum. 
The spontaneously emitted photon has a random direction, so once it is emitted the atom recoils in a random 
direction while the photon does not "return" to the beam. As the spontaneous emission has a random direction 
it does not change, on average, the atomic momentum, as the simulated emission does. In low intensities the spontaneous emission dominates over the stimulated emission so in practice 
the atomic momentum changes only by the effect of stimulated photon absorptions.\cite{cto11}. The size of the momentum "kick" that is given to an atom by a single photon absorption or emission is very small compared 
to the atomic momentum so at first sight it seems strange that we can change the atomic momentum by a considerable 
amount. But the presence of laser ensures a large number of photon absorption and spontaneous emission cycles and the atom gets a net and considerable momentum along the beam propagation direction. It is like to try pushing or stopping a big box by repeatedly hitting it with peebles. 

On average this exchange results to a net transfer of momentum from the light field to the atom and gives rise 
to the so called radiation scattering (or dissipative) force. This force depends on the rate $\Gamma$ with which spontaneous emission occurs and on the momentum carried by the photons. But the mechanical effects of light on atoms
proved to be even more interesting after the discovery of Laguerre-Gaussian (LG) laser beams.\cite{absw92}
The name of this Gaussian beams is after the Laguerre polynomial $L_{p}^{|l|}(x)$ which characterises their amplitude and where $l$ and $p$ are integer indices. The advent of these beams heralded new possibilities in
the manipulation of the atomic gross motion since a LG photon carries not
only a linear momentum but also a quantised orbital angular momentum equal to $l\hbar$ per photon
along the beam propagation axis. The intensity profile of these beams in a direction transverse to the propagation direction is characterised by $p+1$ concentric beams. From this property these beams are also known in the literature
with the nickname "doughnut beams". 

The realisation of such beams in the
laboratory was followed by a sizeable amount of theoretical work on the
mechanical effects of such beams on atoms.\cite{bal13}, \cite{bal18}
The common feature of most of the investigations is that they are limited to cases of
the lowest values of the winding number $l$ and, significantly, they also
ignore modes with non-zero values of the radial mode index $p$. Advances in
techniques for the generation of twisted light have recently enabled modes
with very large radial index, $p$, and/or winding number $l$ to be realised.\cite{schzbl13} 
The current experimental activity on the
production of optical vortices with extremely large values of $l$ and $p$~%
continues.\cite{schzbl13} It has been suggested that such
beams can be exploited in the creation of concentric cylindrical lattices in
which quantum Hall physics with cold atoms can be realised.\cite{lpslldbz15}%

In a recent paper has been clearly pointed out that if the helicity of the LG beam is large enough
and the focussing of the beam very strong then the expression for the scattering force is modified.\cite{lb16} 
The modifications stem from phase gradients
originating from the beam curvature and the Gouy phase most prominently near
the focus plane of the LG light mode. Both the Gouy and the curvature phase
terms have so far been ignored in the analysis, with the Gouy phase strongly dependent on
the values of $l$ and $p$.

\section{The scattering force on a two-level atom}
The prototype physical system under consideration is an atom which is irradiated
by a coherent LG laser beam. In this case and due to the sharply defined frequency of the beam the light can induce electron transitions
only between two atom energy levels, namely a ground state $|1\rangle$ and an excited state $|2\rangle$. These two states have an en energy separation
$E_{12}=\hbar\omega_{0}$, where $\omega_{0}$ is the so called atomic transition frequency. The excited atomic state $|2\rangle$ is characterized by a spontaneous emission rate $\Gamma$.

The LG beam has a frequency $\omega_{L}$ and is considered as propagating along the $z-$direction while is polarised along the $x-$direction.
The electric field of such a beam is given by \cite{bal18}:
\begin{equation}
\mathbf{E}_{l,p}(\mathbf{R})=\frac{1}{2}E_{l,p}(r,z)\exp{\left[i\Theta_{lp}(r,z,\phi)\right]}\exp{(-i\omega_{L} t)}\mathbf{i}+C.C,
\label{elfield}
\end{equation}%
where the amplitude $E_{lp}(r,z)$ is given by
\begin{equation}
E_{l,p}(r,z)=E_{0}f(r,z),
\label{fieldamp}
\end{equation}
with
\begin{equation}
f(r,z)=\frac{1}{\sqrt{1+z^{2}/z_{R}^{2}}}\sqrt{\frac{2p!}{\pi(|l|!+p!)}}\left(\frac{r\sqrt{2}}{w(z)}\right)^{|l|}L_{p}^{|l|}\left(\frac{2r^{2}}{w^{2}(z)}\right)\exp\left(-\frac{r^{2}}{w^{2}(z)}\right),
\label{fterm}
\end{equation}
and the phase $\Theta_{l,p}(r.z,\phi)$ is given by the relation
\begin{equation}
\Theta_{l,p}(r.z,\phi)=kz+ l\phi-(2p+|l|+1) \tan^{-1}(z/z_{R})+\frac{kr^{2}z}{2(z^{2}+z_{R}^{2})}.
\label{fieldphase}
\end{equation}
In the above relations we have that $w(z)=w_{0}\sqrt{1+z^{2}/z_{R}^{2}}$, where $w_{0}$ is the beam waist and $z_{R}=\pi w_{0}^{2}/\lambda$ the Rayleigh range of the beam. The quantity $E_{0}$ is the electric field amplitude which corresponds to a Gaussian beam of the same power and beam waist while $L_{p}^{|l|}$ is the associated Laguerre polynomial. The third and the fourth terms in the RHS of Eq.(4) are the Gouy and the curvature phase respectively. 

The physical basis of the scattering force is the continuous exchange of photons between the laser beam and the atom and their removal from the total system beam+atom to the vacuum due to spontaneous emission. The scattering force for an atom moving with velocity $\mathbf{v}$ is given by :
\begin{equation}
\langle {\bf F}_{{\rm diss}}(\mathbf{v})\rangle=\hbar \Gamma\nabla\Theta_{l,p}\frac{\Omega^{2}/4}{\Delta^{2}(\mathbf{v})+(\Gamma^{2}/4)+(\Omega^{2}/2)}.  
\label{dissipative force}
\end{equation}
In the above relation $\Delta(\mathbf{v})=\omega_{L}-\omega_{0}-(\nabla\Theta\cdot\mathbf{v})$ is the detuning, i.e the difference between the laser frequency and the atomic transition frequency. The quantity $\nabla\Theta\cdot\mathbf{v}$ is the Doppler shift of the laser beam frequency as observed by the moving atom \cite{abp94}. The quantity $\Omega$ is the so called Rabi frequency, i.e. the frequency with which occur the transitions induced by the laser namely the stimulated photon emission and absorption. The Rabi frequency is given by $\Omega=\Omega_{0}f(r,z)
$ and is associated with the laser power and intensity since $\Omega_{0}=\sqrt{I_{0}\Gamma/2I_{s}}$
where $I_{S}$ is the saturation intensity for the corresponding atomic transition.\cite{mvdS99}
The formula in Eq.(\ref{dissipative force}) has the following meaning: the force results from scattering of photons away from the beam so it depends on the momentum of each photon $\hbar\nabla\Theta$ (if the light field is a plane wave then the phase $\Theta$ for a plane is just $kz$, thus $\nabla\Theta=k$) and on the rate $\Gamma$ at which this scattering occurs. The spontaneous emissions have a statistical character which is reflected at the Lorentzian term $\frac{\Omega^{2}/4}{\Delta^{2}(\mathbf{v})+(\Gamma^{2}/4)+(\Omega^{2}/4)}$. This term reaches its maximum value when the Doppler shift compensates the detuning, i.e. when $(\nabla\Theta\cdot\mathbf{v})=\omega_{L}-\omega_{0}$ and thus the interaction between the beam and the atom is resonant .

If we consider that the beam is an LG beam with large helicity the dissipative force for an atom near the beam focus, $z=0$, is given by:
\begin{equation}
\langle {\bf F}_{{\rm diss}}(\mathbf{v})\rangle=\hbar \Gamma\frac{\Omega^{2}/4}{\Delta^{2}(\mathbf{v})+(\Gamma^{2}/4)+(\Omega^{2}/2)}\left[k\left( 1-\frac{(2p+|l|+1)}{kz_{R}}+\frac{r^{2}}{2z_{R}^{2}}\right)\widehat{{\bf z}}+\frac{l}{r}\widehat{{\bf \phi }}\right]{.}
\label{dissforce}
\end{equation}
As we see the force is made up from two components one along the axial direction and one along the azimuthal direction.
The later as has been shown is responsible for a torque along the axial direction on the atom \cite{bpa94}. We are going to focus on the axial component.

We consider the case where the radial index $p=0$ and we assume that our atom is located at a 
a radial distance equal to $r=w_{0}\sqrt{|l|/2}$ , i.e. at the radial distance at which the beam intensity has its maximum 
when $p=0$ and thus atom-beam interaction is stronger. We also assume a very large value of the index $l$ such that $|l|>>1$. In this we can show that the dissipative force assumes the form:
\begin{equation}
\langle {\bf F}_{{\rm diss}}(\mathbf{v})\rangle_{z}=\hbar \Gamma k\frac{\Omega^{2}/4}{\Delta^{2}(\mathbf{v})+(\Gamma^{2}/4)+(\Omega^{2}/2)}\left( 1-\frac{|l|}{2kz_{R}}\right)\widehat{{\bf z}}.
\label{dissforceapprox}
\end{equation}
Now as we know the atomic velocity is far smaller than the speed of light so a power series expansion around $v=0$ of the force gives us the result:
\begin{equation}
\langle {\bf F}_{{\rm diss}}(\mathbf{v})\rangle_{z}=\langle {\bf F}_{{\rm diss}}(\mathbf{0})\rangle_{z}+av+... ,
\label{powerapprox}
\end{equation}
where $\langle {\bf F}_{{\rm diss}}(\mathbf{0})\rangle_{z}$ is the force for zero atomic velocity and the coefficient $a$ is given by:
\begin{equation}
a=\hbar k^{2}\left(1-\frac{|l|}{2kz_{R}}\right)^{2}\frac{s}{(1+s)^{2}}\frac{\Delta(0)\Gamma}{\Delta^{2}(0)+(\Gamma^{2}/4)}.
\label{acoeff}
\end{equation}
with $s$ being the so called saturation parameter given by $s=(\Omega^{2}/2)/(\Delta^{2}(0)+\Gamma^{2}/4)$. The coefficient $a$ is negative when the detuning $\Delta(0)$ is negative. In this case the last term in Eq.(\ref{powerapprox}) is a damping term. After all the above analysis we can write, for negative detuning, the total force as:
\begin{equation}
\langle {\bf F}_{{\rm diss}}(\mathbf{v})\rangle_{z}=\hbar\Gamma k\frac{\Omega^{2}/4}{\Delta^{2}(0)+(\Gamma^{2}/4)+(\Omega^{2}/2)}\left(1-\frac{|l|}{2kz_{R}}\right)\widehat{{\bf z}}-|a|v\widehat{{\bf z}}.
\label{3partforce}
\end{equation}
As we clearly see the total force force expression is made up from three terms: the first term $F_{o}=\hbar \Gamma k\frac{\Omega^{2}/4}{\Delta^{2}(0)+(\Gamma^{2}/4)+(\Omega^{2}/2)}$ is the well-known expression for the axial scattering force, the second term $F_{l}=\frac{\hbar \Gamma |l|}{2z_{R}}\frac{\Omega^{2}/4}{\Delta^{2}(0)+(\Gamma^{2}/4)+(\Omega^{2}/2)}$ is opposite to the first and is "quantised" since $|l|=1, 2, ...$. The second term of the force has a considerable amount for large values of the beam helicity $l$ and for tight focussing i.e. for small values of beam waist $w_{0}$ and Rayleigh range $z_{R}$. The third term $F_{D}=-|a|v$ is a damping force. This term is responsible for the optical molasses effect, a viscous type of motion of a sample of atoms when they are irradiated by counter-propagating laser beams.\cite{chbca85}  

\section{Connection with the Millikan experiment}
\begin{figure}[h!]
\centering
\includegraphics[width=6in]{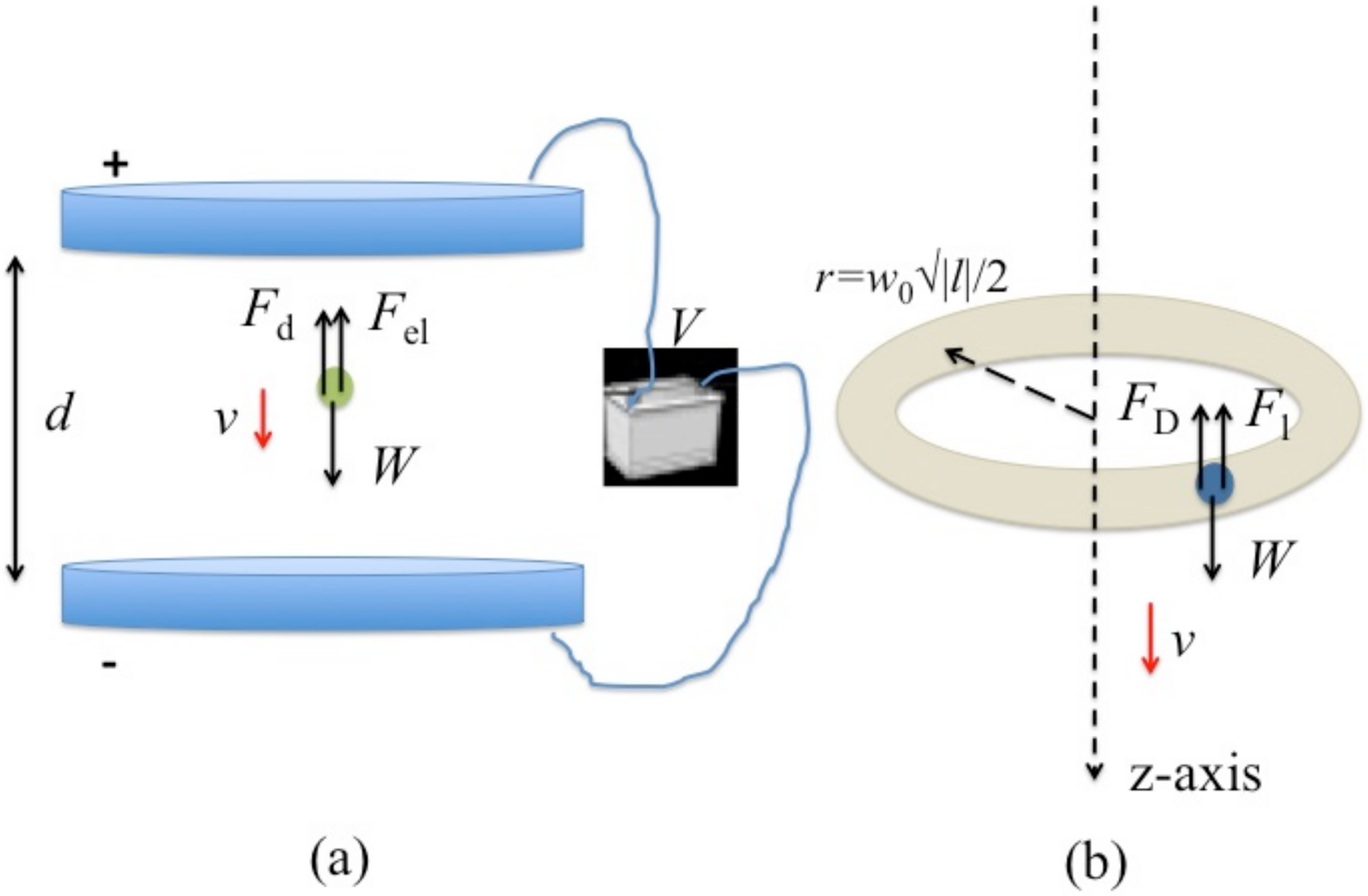}
\caption{The involved forces in the Millikan experiment (a) and in the case of the interaction of the atom with a highly twisted LG beam (b). The latex sphere is negatively charged.}
\label{forcesplot}
\end{figure}

As we know in the famous Millikan experiment a small charged dielectric object, like an oil droplet or a latex sphere,
experiences three forces: its weight $W=mg$, the electric force from a uniform electric field $E$ given by  $F_{{\rm el}}=Eq$ which is quantised since the charge of the droplet is an integer multiple of the electron charge $q=n|e|$ and finally a drag force due to air which is opposite to the velocity of the charged particle $F_{d}=-K_{d}v$.  We consider that the weight direction is along the positive $z-$axis direction. The experimental set up is arranged in such a way that the electric force is in the opposite direction of the weight. In Fig.\ref{forcesplot} we show the involved forces in the two cases which we study in this paper. Thus we can say that the resultant force on the particle is given by:
\begin{equation}
\langle {\bf F}_{res}\rangle _{z}=(mg-n|e|E-K_{d}v)%
\widehat{{\bf z}}.
\label{resforce}
\end{equation}
If we compare the expressions in Eq.(12) and Eq.(11) we see a clear analogy. The force $F_{o}$ is the analogue 
of the weight $W$, the force $F_{l}$ is the analogue of $F_{{\rm el}}$ while in both expressions we have damping forces opposite to the
atom's and particle's velocities respectively.
The similarity of the two cases is even more striking. We consider a typical undergraduate Millikan
experiment with latex spheres of density $\rho=1.05 g/cm^{3}$ and
diameter $D=1.070 \mu m$.\cite{colorado} The spheres are considered to be inside a uniform electric field
created by two plates at a distance $d=0.6 mm$ apart and kept at potential
difference $V=50V$. The charge of the sphere in the experiment varied from 1
to 10 times the charge of electron. In this case the weight
of the sphere is $W=5.3\times 10^{-14} N$ while the electric force, if we
consider a charge $n=2|e|$, is $F_{{\rm el}}=2.7\times 10^{-15} N$. The
ratio of the two forces is $F_{{\rm el}}/W=0.05$. Assume now that we
irradiate a Cs atom with a beam of wavelength $\lambda=852.35$ nm which
excites the transition $6^{2}S_{1/2}-6^{2}S_{3/2}$. The beam is of LG type
with a waist $w_{0}=7 \mu m$ and mode indices $p=0$ and $l=137$. The value of the spontaneous emission rate
of the excited state is $\Gamma=3.25\times10^{7}$ r/s, the detuning has the value $\Delta(0)=0.5\Gamma$
while the Rabi frequency is $\Omega_{0}=4\Gamma$. In this
case the constant part of the force in eq.(\ref{dissforceapprox}) which
is $F_{o}=1.19\times10^{-20}$ N,
while the second part is $F_{%
{l}}=5.96\times10^{-22}$ N. The ratio of these two forces is $F_{%
{l}}/F_{0}=0.05$. This is in an impressive agreement
with the ratio of the electric force to the weight. 

The above numerical example shows that the axial dynamics of an atom inside 
a highly structured LG light beam "mimics" the behaviour of a
charged sphere inside a uniform electric field. Generalising we could say
that the atom mimics a sphere of "mass" $\hbar k\Gamma/g$ of "charge" $%
|l||e|$ inside an "electric field" $\hbar\Gamma/(z_{R}|e|)$.

But this analogy is even more striking. Now let us consider the strengths of the damping forces.
The  coefficient $K_{d}$ of the drag force on the latex sphereis given by $K_{d}=6\pi\eta r/(1+b/(pr))$
where $\eta$ is the viscosity of the air, $r$ the radius of the charged latex sphere, $b$ is an experimentally determined 
constant and $p$ the atmospheric pressure. For the parameters we used in our numerical work the value of the coefficient 
$K_{d}=1.6\times 10^{-10}$ kg/s. In this case the latex spheres acquire a terminal velocity equal to $v_{t}=0.3$ mm/s. This gives us 
an estimation of the order of magnitude of the damping force which is $F_{d}=5\times 10^{-14}$ N. Comparing this force to the weight of the latex sphere we have $F_{d}/W=0.95$. In the case of the damping force acting on the atom
we can achieve such a value for the ratio $F_{D}/F_{o}$ for an atomic speed of $v=40$ m/s which is a speed
that can be achieved by cooling the atomic motion from an initial thermal speed (of the order $10^{3}$ m/s). As we know
the lower speed that can be achieved with scattering forces acting on a Cs atom is $0.088$ m/s. All the above numerical examples show clearly that the analogy between these physically different effects is complete and justified.

\section{Conclusion}

We presented the analogy which exists between the axial 
scattering force exerted on a two-level atom when it is irradiated
by a highly twisted LG beam, and the resultant of the forces exerted on
a dielectric charged particle inside a uniform electric field. 
We would like to emphasize the term "analogy" because in the case
of the Millikan experiment we have three distinct and fundamentally 
different forces namely the weight, the electrostatic force and the buoyant force, while
in the case of the irradiation of the atom by a twisted beam 
the "quantized" force $F_{l}$ and the damping force $F_{D}$  forces were just correction terms of the scattering force:
$F_{l}$ becomes important when the beam helicity is large and its focussing is tight, while the damping force $F_{D}$
appears as the atomic speed is such that Doppler shifts the laser beam frequency.
The most provoking similarity is the "quantization" in the electric force and the scattering force term $F_{l}$.
We must point out that the term $F_{l}$ emerges when we use highly twisted and strongly focussed LG beams.
If we had used a typical Gaussian laser beam this term would not have appeared. On the contrary the quantization of the
electric forces is a result of a fundamental property of the electric charge. We must also point out that the additional term in the force component along the $z$-direction
which will result in the force $F_{l}$ appears because we have kept in our workings the Gouy phase and the curvature phase terms respectively.
The Gouy phase has been shown to originate from the in-plane confinement of the focussed beam \cite{fw01},
or as a geometrical quantum effect,  a result of the uncertainty principle when the beam is transmitted through a space with a modified volume \cite{hr96}.
These Gouy phase is significant when the winding numbers are high and also for tight focussing. Tight focussing is a problem in the sense that
the paraxial approximation breaks down and light can not represented by the LG modes.

Another interesting point is the role played by the other characteristic number of the LG beam, namely the radial index $p$, which in the above 
calculations was taken as equal to zero in order to simplify the calculations. This index is also an integer and could, thus, contribute to the "quantized" force $F_{l}$ even in the case where
we have a beam with $l=0$ but a non zero $p$. As it has been pointed out  researchers have not payed much attention to this index and thus there are a few works concerning
its physical significance \cite{plrz13}, \cite{pk15}. As it has been pointed out this number can be considered as as a conjugate quantity of the spatial confinement of the beam \cite{plrz13}. The fact that this number
appears in the expressions for the radiation pressure forces on atoms is a nice chance for observing its physical effects in an experiment.
The cylindrical symmetry of our modes and their spatial confinement are at the heart of the results taken in this paper.

\begin{acknowledgments}

This project was supported by King Saud University, Deanship of Scientific Research, College of Sciences Research Center.

\end{acknowledgments}

\end{document}